\documentclass[a4paper,twocolumn]{esapub2005} % European paper
\usepackage{times}
\usepackage{natbib}
\usepackage{graphicx}
\usepackage{psfig}

\title{Swift/XRT follow-up observations of INTEGRAL AGNs}
\author{R. Landi, A. Malizia, L.Bassani, N. Masetti, J.B. Stephen, F.Gianotti, F. Schiavone}
\affil{IASF-INAF, Bologna, Italy}
\author{A. Bazzano, P. Ubertini}
\affil{IASF- INAF, Roma, Italy}
\author{A. J. Bird, A. J. Dean}
\affil{University of Southampton, Southampton, UK}
\author{R. Walter}
\affil{INTEGRAL Science Data Centre, Versoix, Switzerland}

\begin{document}


\maketitle

\begin{abstract}
  In five years of operation, data from INTEGRAL has been used to discover a large
  number of gamma-ray sources, a substantial fraction of which have turned out to
  be active galactic nuclei (AGN). Recently Bassani et al. (2006) \citep{ba06} have
  presented a sample of around 60 AGNs of which some still lack
  optical identification or information in the X-ray band. In this
  work we present X-ray data for 8 objects acquired with the XRT
  telescope on-board Swift satellite. The XRT positioning has allowed
  us to identify the optical counterparts and to classify their types
  through follow-up measurements. Analysis of these data has also
  provided information on their spectra below 10 keV.
\end{abstract}

\section{Introduction}
Our knowledge of the hard X-ray sky has been widened thanks to the
capabilities of IBIS \citep{ub03}, the gamma-ray imager on board the
INTEGRAL satellite \citep{wi03}. Since its launch in October 2002,
INTEGRAL has surveyed a large fraction of the sky above 20 keV at a
mCrab sensitivity level with a typical localization accuracy of 2-3'
\citep{bi06}.  The number of hard X-ray sources detected in the
second IBIS survey has increased by 70\% compared to the first one for
a total of 209 objects \citep{bi04,bi06}.  The main category
($\sim$50\% of the sample) is still represented by galactic accreting
binaries (HMXB and LMXB), although there has been a five-fold increase
in AGN detections over the 1st catalogue while the number of
unclassified objects has doubled as a direct consequence of a
wider and deeper sky coverage.  Masetti and co-authors
\citep{p1,p2,p3,p4,p5} are currently executing a campaign with the goal of
identifying the still unknown sources through optical
spectroscopy. Their results indicate that half of the unidentified
INTEGRAL sources turn out to be optically classified as nearby AGNs.\\
The INTEGRAL error box associated with the hard X-ray sources (2-3')
is often too large (and therefore too crowded) to allow an optical follow-up
observation.  Therefore we have searched in the \emph{Swift} X-Ray
Telescope (XRT) \citep{ge04} archive for X-ray observations of newly
detected IBIS sources in order to locate them with arcsec accuracy;
this us allowed their optical classification in many cases 
(see \citep{p1,p2,p3,p4,p5}).\\
In this work we present X-ray data (0.2--10 keV) for a set of 8
identified with AGN.

\begin{table*}[th!]
  \begin{center}
    \caption{Spectral analysis of the sample} \vspace{1em}
    \renewcommand{\arraystretch}{1.2}
    \begin{tabular}[h]{llcccc}
      \hline
      Source    & Type    &  N$_{HGal}$             & N$_{H}$               &  $\Gamma$  & Flux (2-10 keV)   \\
                &         & (10$^{21}$ cm$^{-2}$)   & (10$^{22}$ cm$^{-2}$)  &           & (10$^{-12}$ erg cm$^{-2}$ s$^{-1}$\\
      \hline
IGR J07597-3842$^{\star}$ & Sey 1.2     & 0.630  & $<$0.05 &                 1.80$^{+0.05}_{-0.03}$  & 22.0 \\
IGR J12415-5750$^{\star}$ & Set 2       & 0.346  & $<$0.11 &                 1.70$^{+0.13}_{-0.11}$  & 8.1  \\
IGR J14492-5535 & unclas.    & 0.500  & 10.1$^{+6.3}_{-4.3}$     & 1.8$^{FIX}$             & 2.1  \\
IGR J16482-3036$^{\star}$ & Sey 1       & 0.176  & 0.13$^{+0.05}_{-0.13}$   & 1.71$^{+0.11}_{-0.09}$  & 10.0 \\
IGR J16558-5203 & Sey 1.2     & 0.304  & $<$0.011                 & 1.85$^{+0.06}_{-0.04}$  & 18.0 \\
IGR J17488-3253 & Sey 1 ?     & 0.530  & 0.22$^{+0.07}_{-0.05}$   & 1.60$^{+0.13}_{-0.10}$  & 14.0 \\
IGR J20187+4041 & unclas.    & 11.9   & 19.3$^{+20.9}_{-11.6}$   & 1.8$^{FIX}$             & 1.8  \\
IGR J20286+2544 & SB/Sey 2 & 0.261  & 42.3$^{+19.5}_{-28.5}$   & 1.8$^{FIX}$             & 2.3  \\
      \hline \\
      \end{tabular}
  \end{center}
\end{table*}

\section{Spectral Analysis}
The XRT data reduction was performed using the XRTDAS v. 2.4 standard data
pipeline package ({\sc xrtpipeline} v. 0.10.3), in order to produce
screened event files. All data are extracted only in the Photon Counting
(PC) mode \citep{hi04}, adopting the standard grade filtering (0--12
for PC) according to the XRT nomenclature. Events for spectral analysis
were extracted within a circular region of radius 20$^{\prime \prime}$,
which encloses about 90\% of the PSF at 1.5 keV \citep{mo04}
centered on the source position.
The background was extracted from various source-free regions close to the
X-ray object of interest using both circular and annular regions of
various radii, in order to ensure an evenly sampled background. In all
cases, the spectra were extracted from the corresponding event files using
{\sc XSELECT} software and binned using {\sc grppha} in a manner so 
that the $\chi^{2}$ statistic could reliably be used. We used the
latest version (v.008) of the response matrices and created individual
ancillary response files (ARF) using {\sc xrtmkarf}. Spectral analyses
were performed using XSPEC version 12.2.1.\\
In figures  1 to 8 we show  the XRT 0.3-10 keV images for
the eight new AGNs discovered by INTEGRAL together with their relative X-ray spectra. 
In each XRT field of view the INTEGRAL error (black circle) is shown.
For those sources with more than one pointing (see label star in the table), we performed the spectral
analysis of each observation individually in order to search for spectral variability, and then
we analyzed the combined spectra, sa as to improve the statistical
quality of the data. For this prelinary analysis, due to the limited spectral signal often available,
we employed a simple power law absorbed by both a Galactic \citep{dl90} and an intrinsic column density.
Due to the low statistical quality of the data, in the case of IGR J14492-5535, IGR J20187+4041
and IGR J20286+2544 we fixed the photon index to a canonical AGN value ($\Gamma$=1.8).
This baseline model provides a quite good fit to the X-ray data  for all the 
AGNs herein analyzed and the spectral parameters obtained for
each source are reported in table 1.

\begin{figure*}[th!]
\begin{center}
\hspace{.1cm}
\hspace{-.8cm}
%\centering{
\mbox{\psfig{file=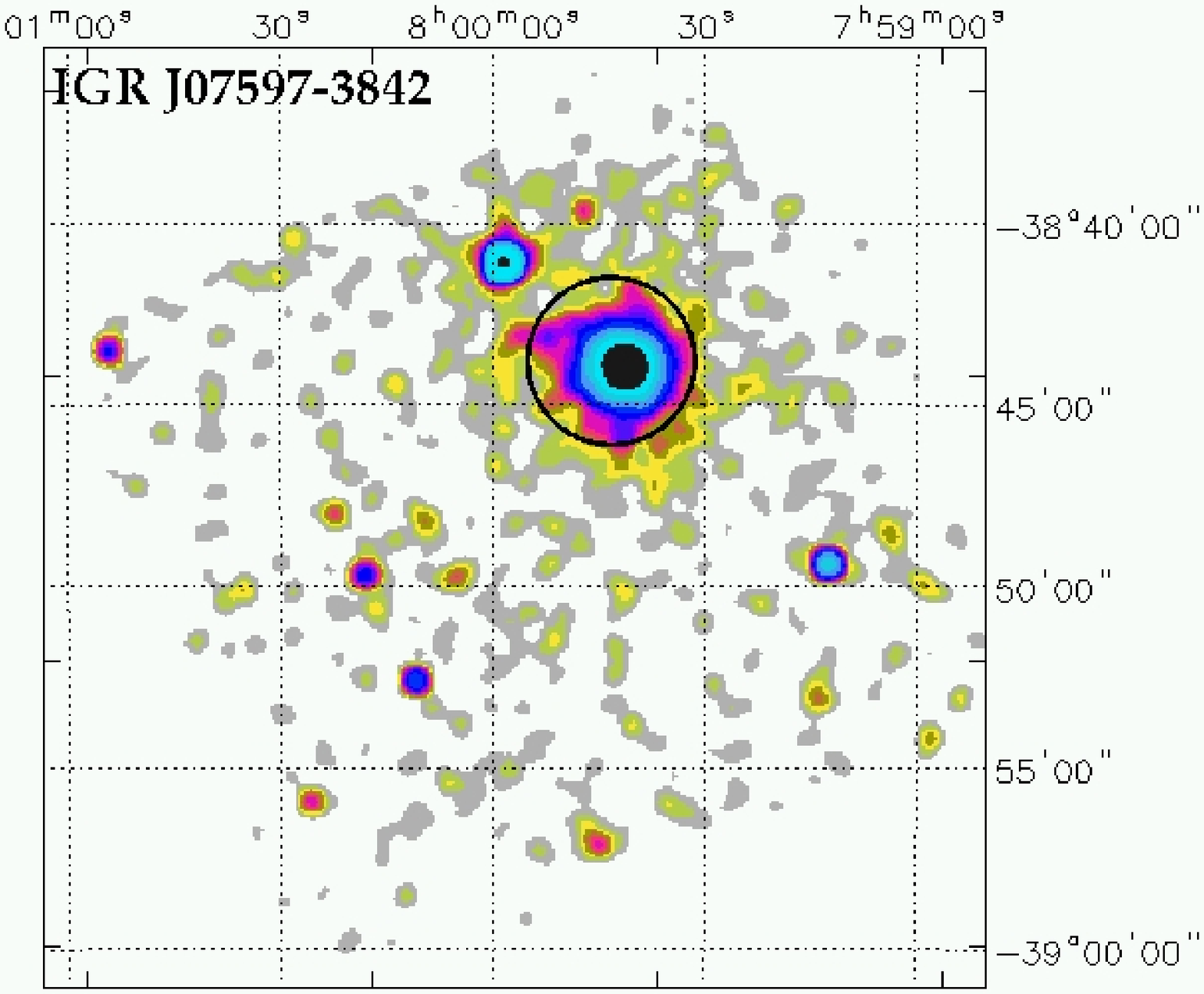,width=5.5cm}}%}
\hspace{-1.0cm}
%\centering{
\mbox{\psfig{file=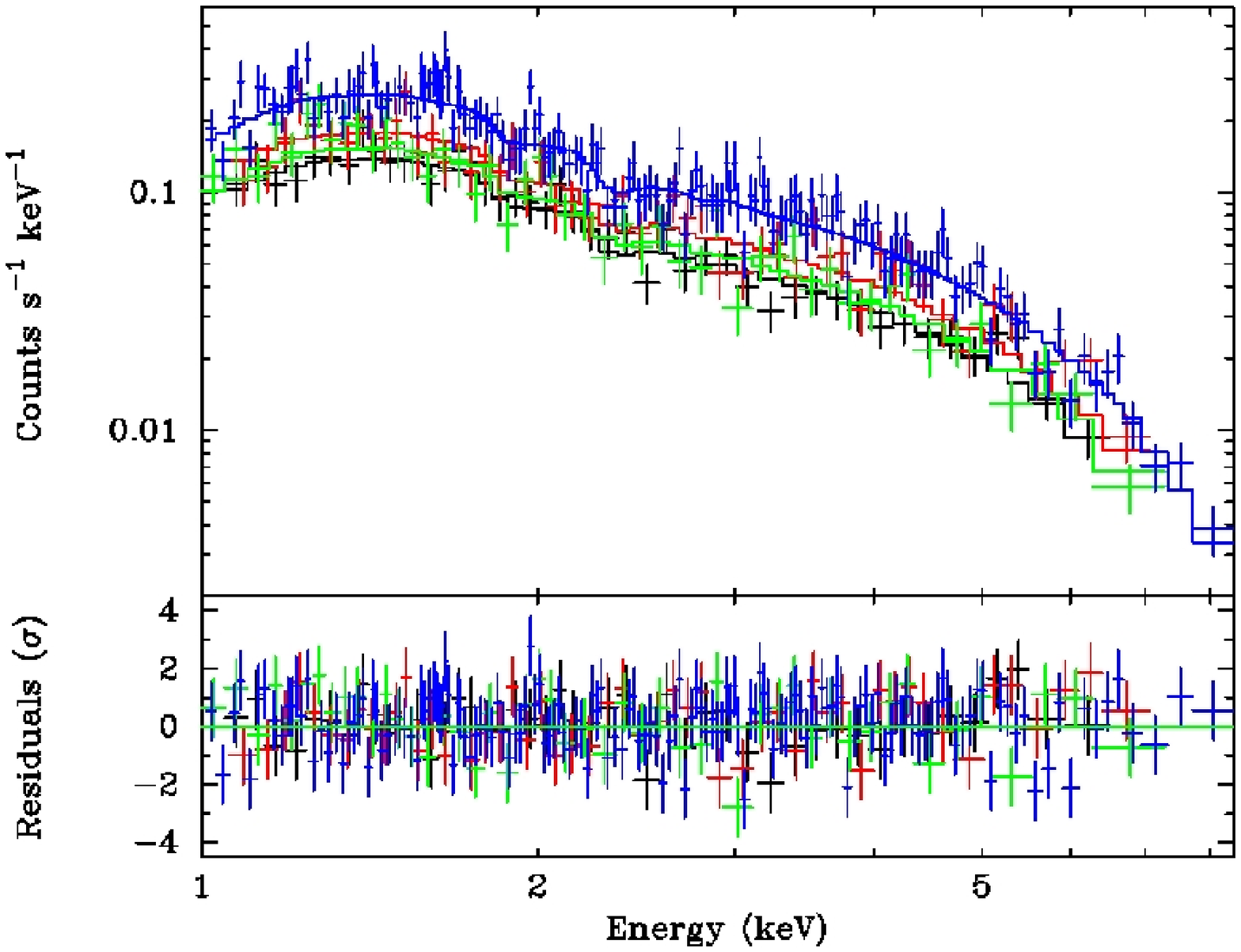,width=5.5cm}}%} 
\end{center}
\end{figure*}

\begin{figure*}[th!]
\begin{center}
\hspace{.1cm}
\hspace{-.8cm}
%\centering{
\mbox{\psfig{file=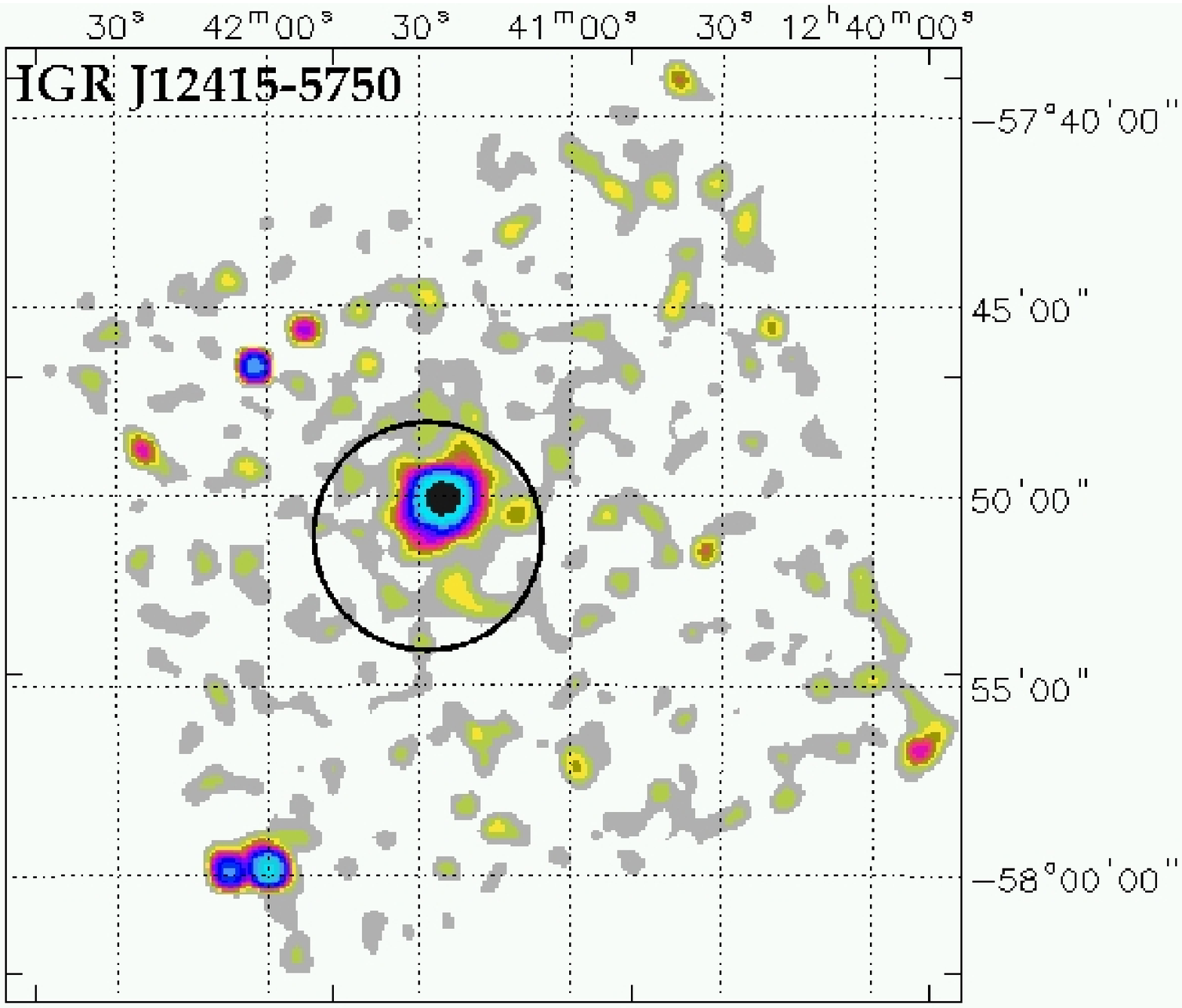,width=5.5cm}}%}
\hspace{-1.0cm}
%\centering{
\mbox{\psfig{file=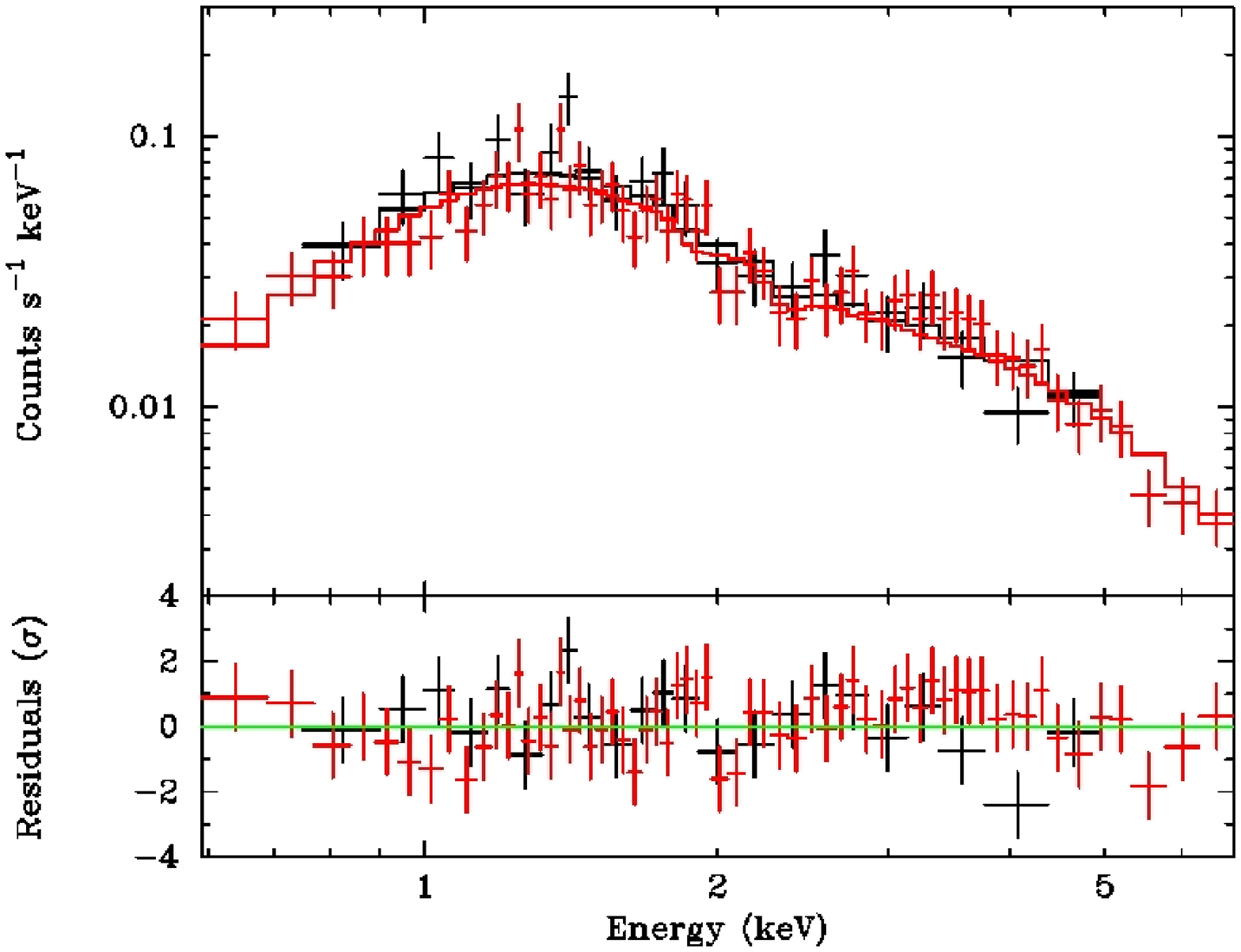,width=5.5cm}}%} 
\end{center}
\end{figure*} 

\begin{figure*}[th!]
\begin{center}
\hspace{.1cm}
\hspace{-.8cm}
%\centering{
\mbox{\psfig{file=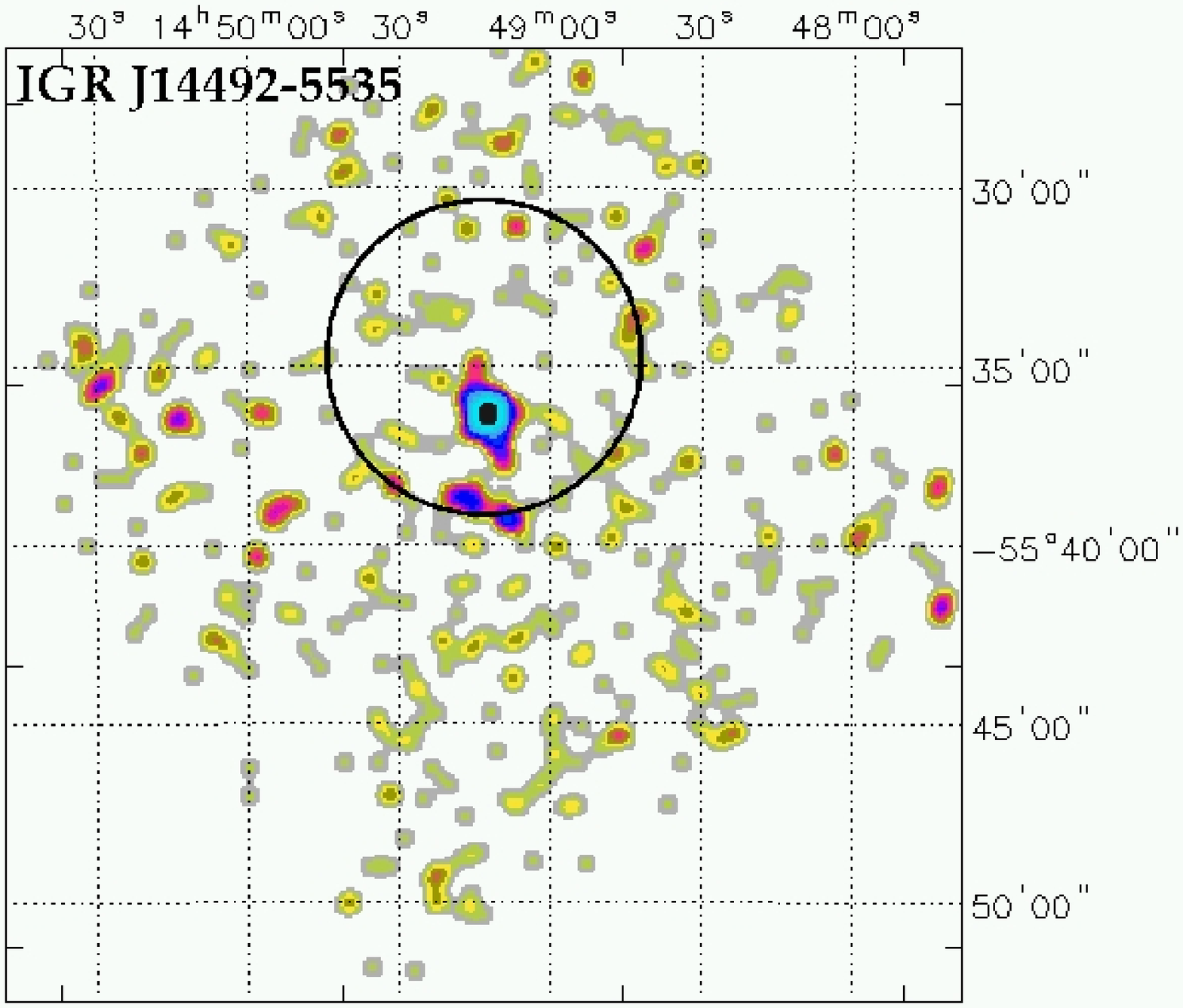,width=5.5cm}}%}
\hspace{-1.0cm}
%\centering{
\mbox{\psfig{file=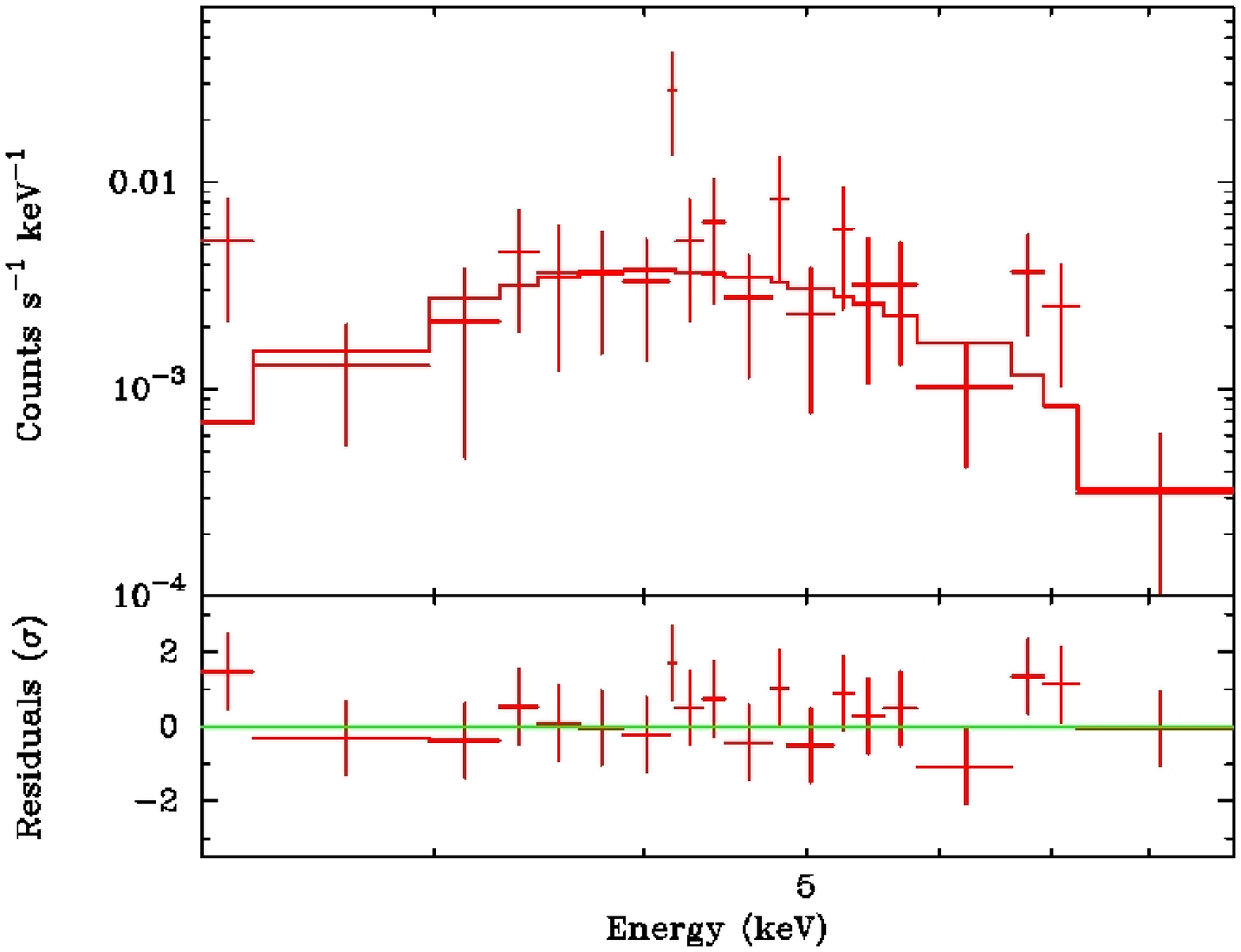,width=5.5cm}}%} 
\end{center}
\end{figure*} 

\begin{figure*}[th!]
\begin{center}
\hspace{.1cm}
\hspace{-.8cm}
%\centering{
\mbox{\psfig{file=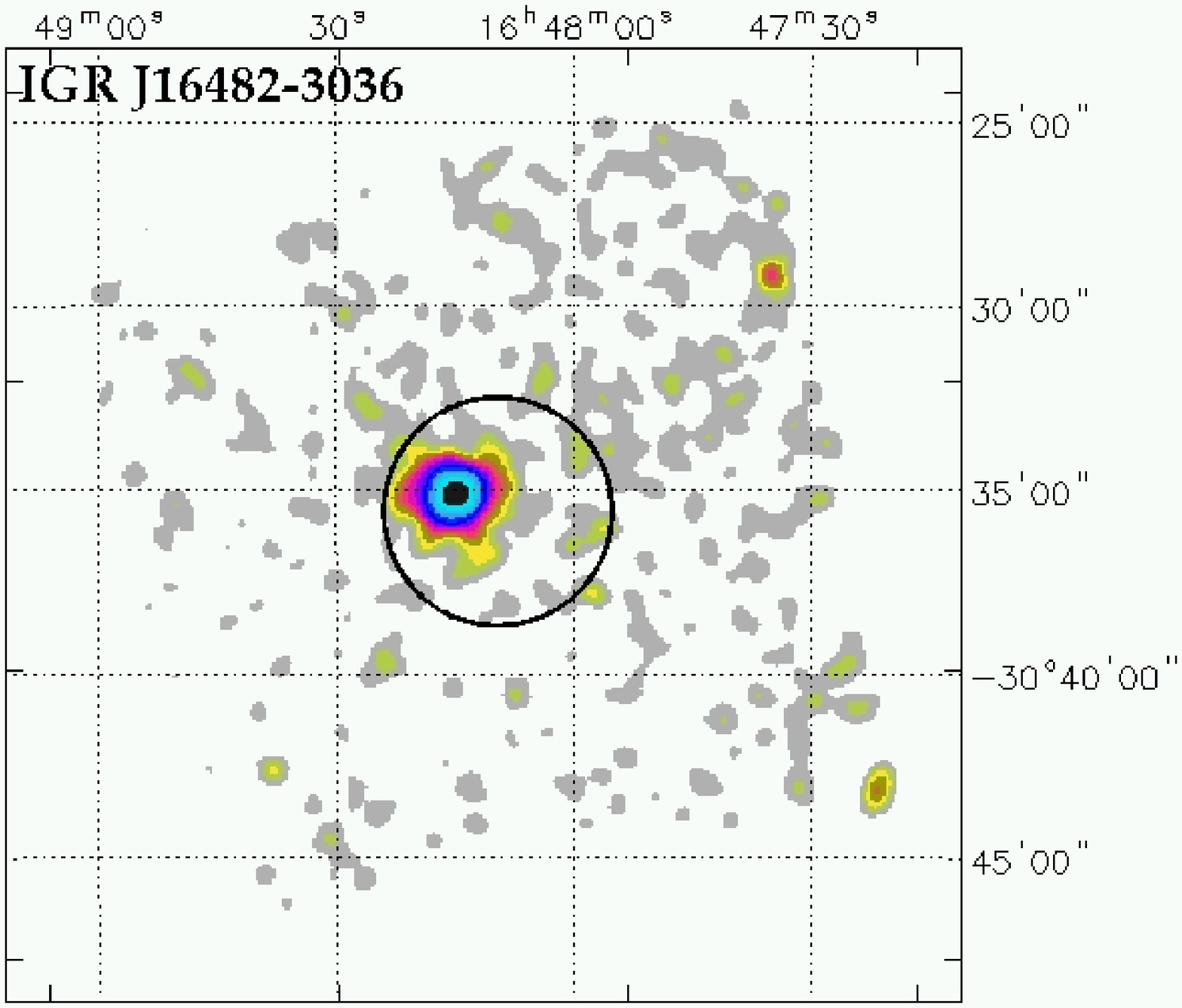,width=5.5cm}}%}
\hspace{-1.0cm}
%\centering{
\mbox{\psfig{file=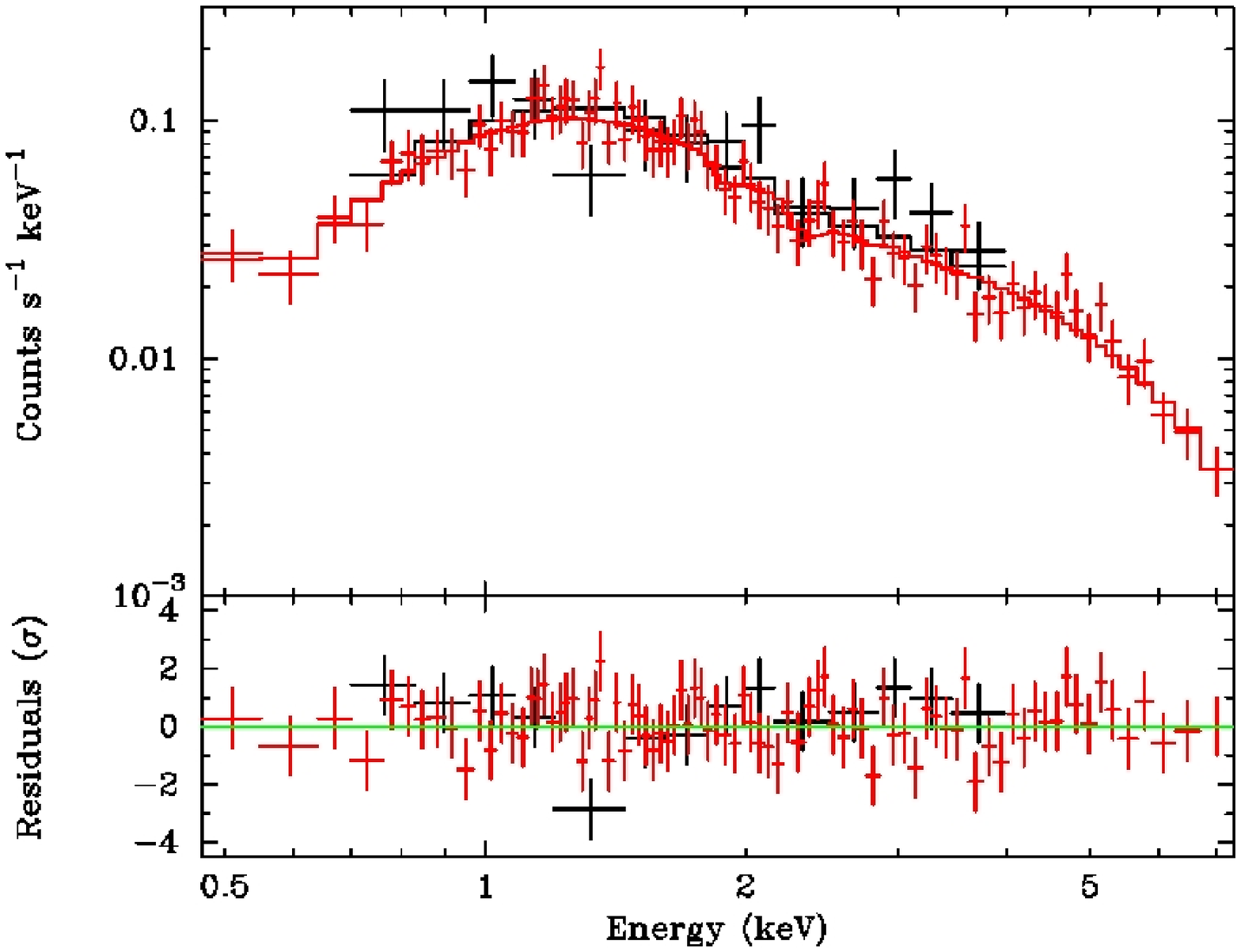,width=5.5cm}}%} 
\end{center}
\end{figure*} 

\begin{figure*}[th!]
\begin{center}
\hspace{.1cm}
\hspace{-.8cm}
%\centering{
\mbox{\psfig{file=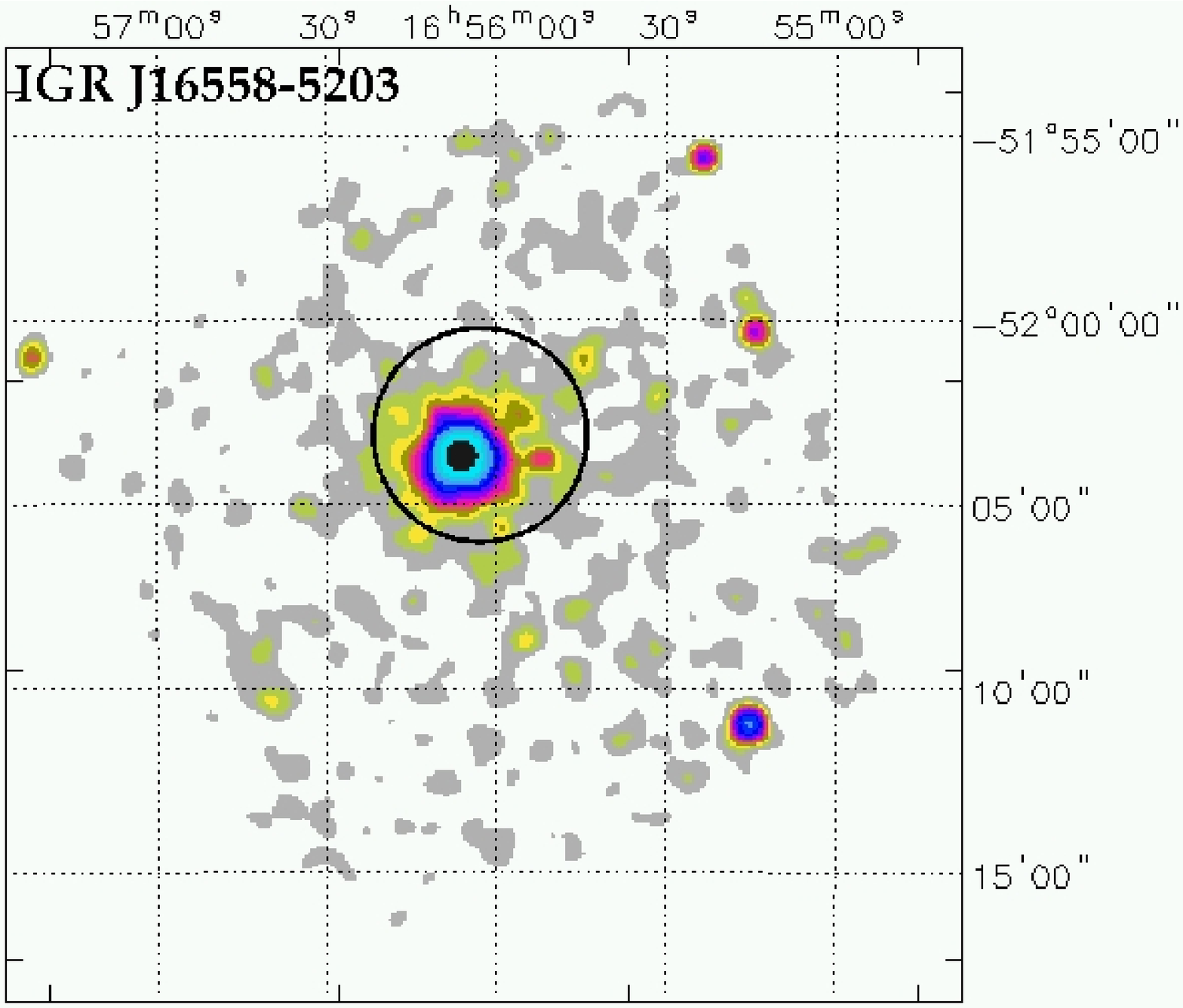,width=5.5cm}}%}
\hspace{-1.0cm}
%\centering{
\mbox{\psfig{file=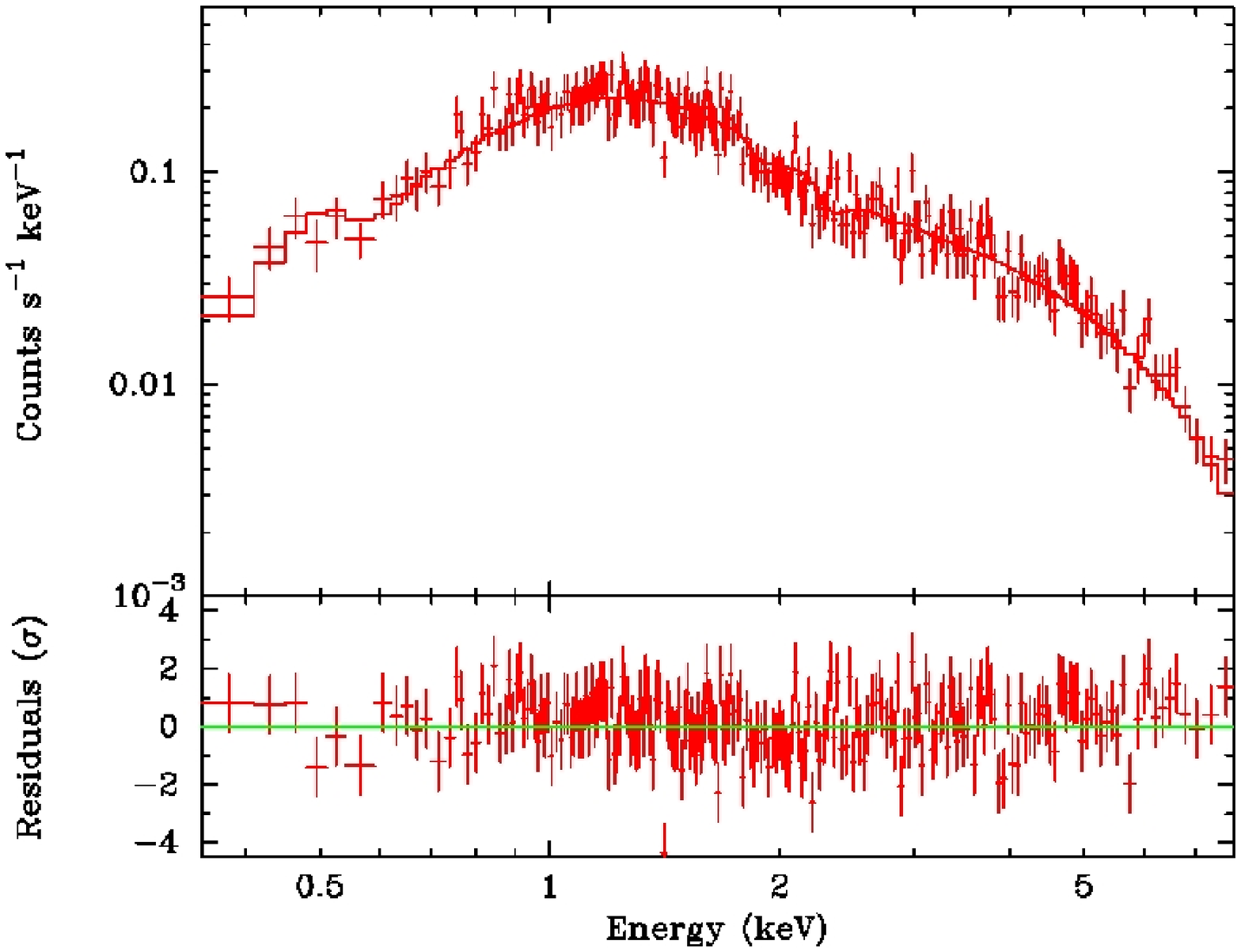,width=5.5cm}}%} 
\end{center}
\end{figure*} 

\begin{figure*}[th!]
\begin{center}
\hspace{.1cm}
\hspace{-.8cm}t
%\centering{
\mbox{\psfig{file=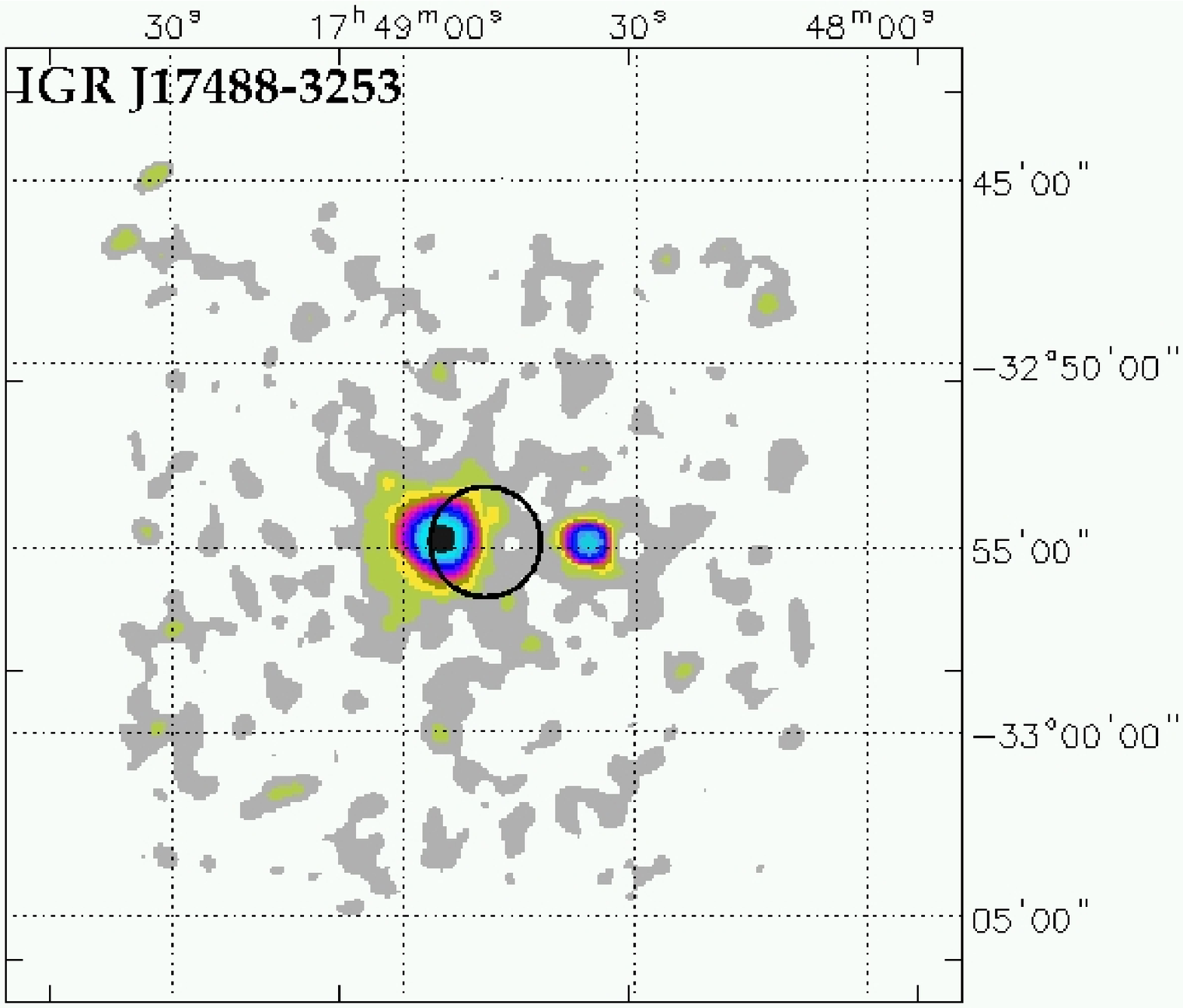,width=5.5cm}}%}
\hspace{-1.0cm}
%\centering{
\mbox{\psfig{file=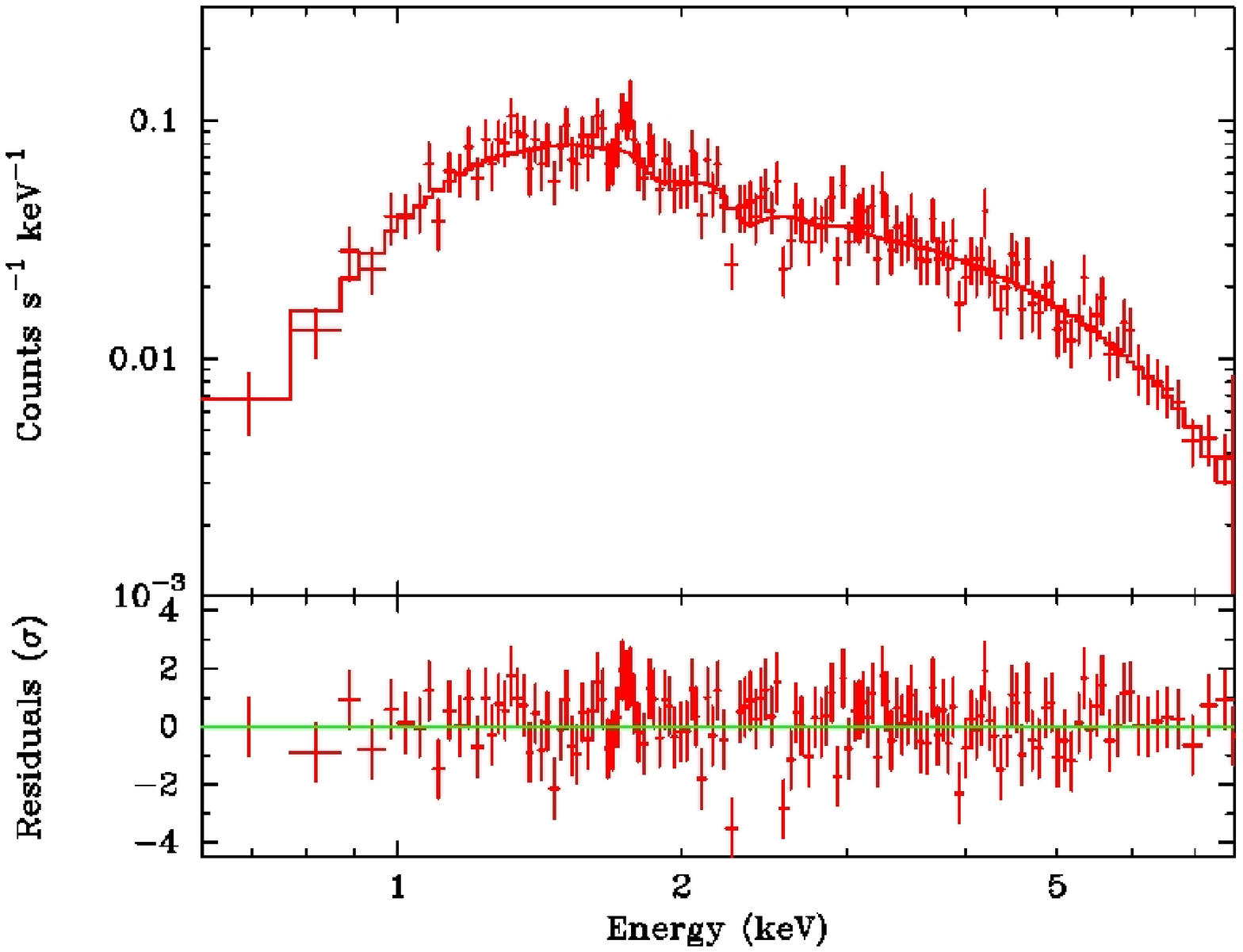,width=5.5cm}}%} 
\end{center}
\end{figure*} 

\begin{figure*}[th!]
\begin{center}
\hspace{.1cm}
\hspace{-.8cm}
%\centering{
\mbox{\psfig{file=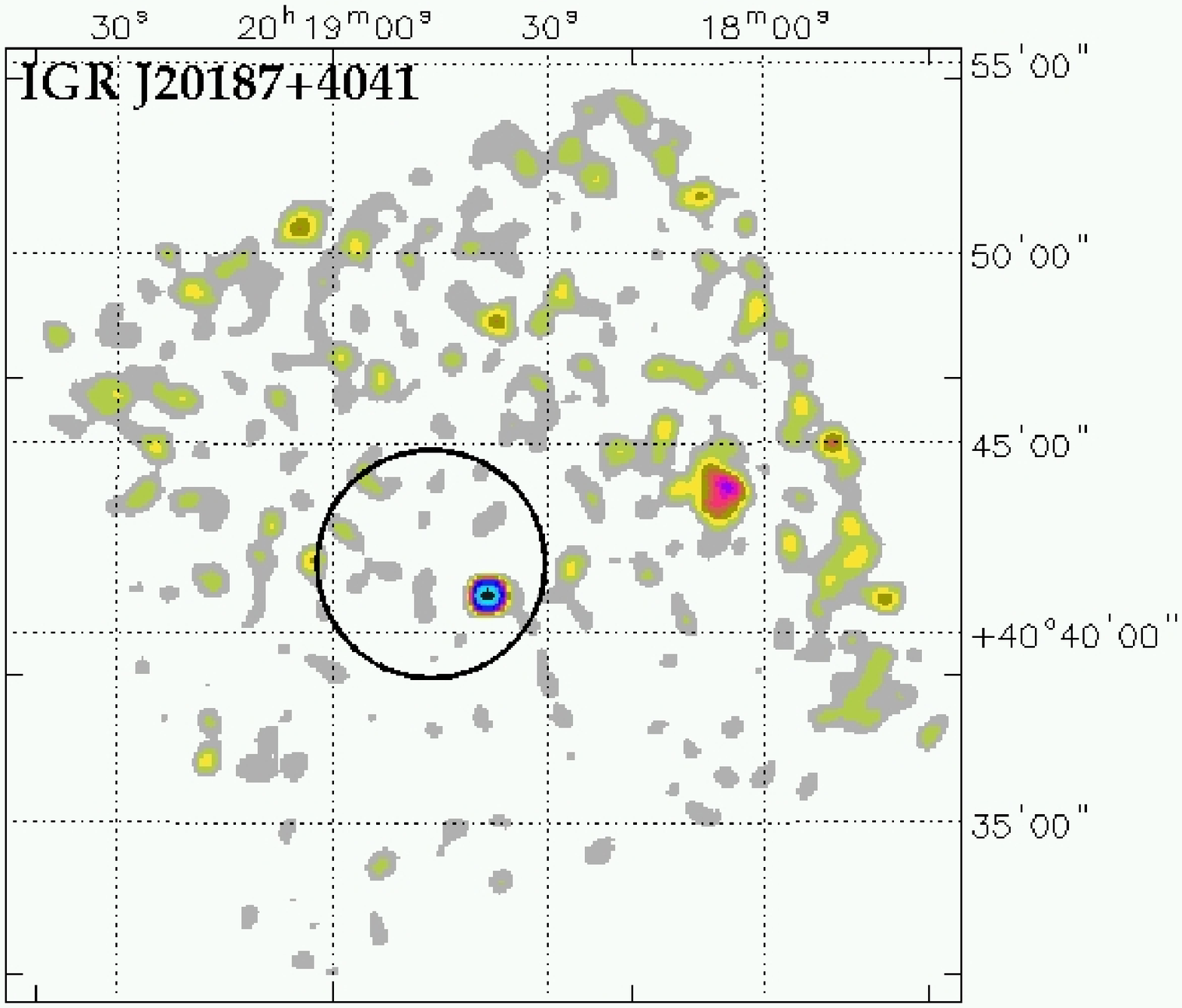,width=5.5cm}}%}
\hspace{-1.0cm}
%\centering{
\mbox{\psfig{file=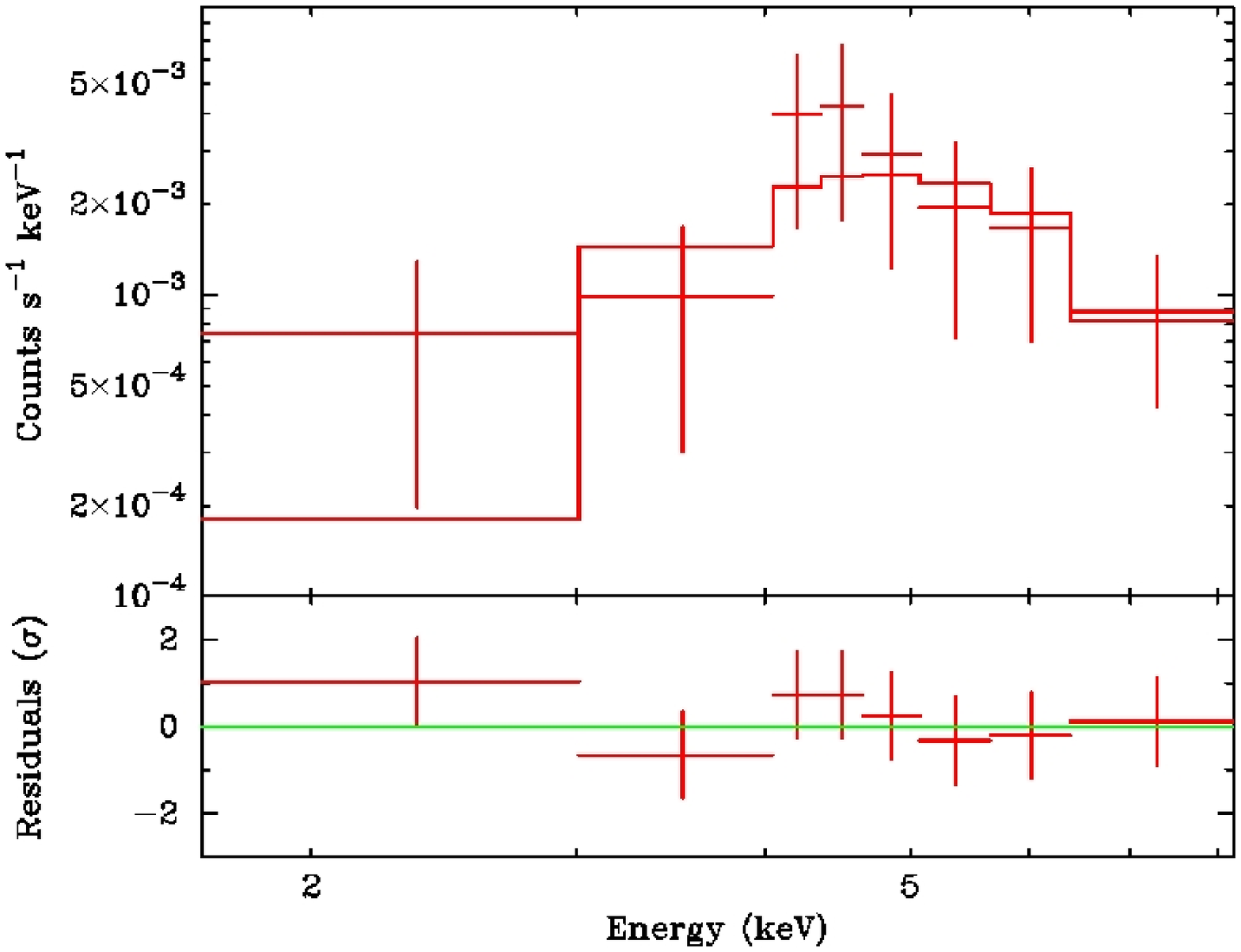,width=5.5cm}}%} 
\end{center}
\end{figure*} 

\begin{figure*}[th!]
\begin{center}
\hspace{.1cm}
\hspace{-.8cm}
%\centering{
\mbox{\psfig{file=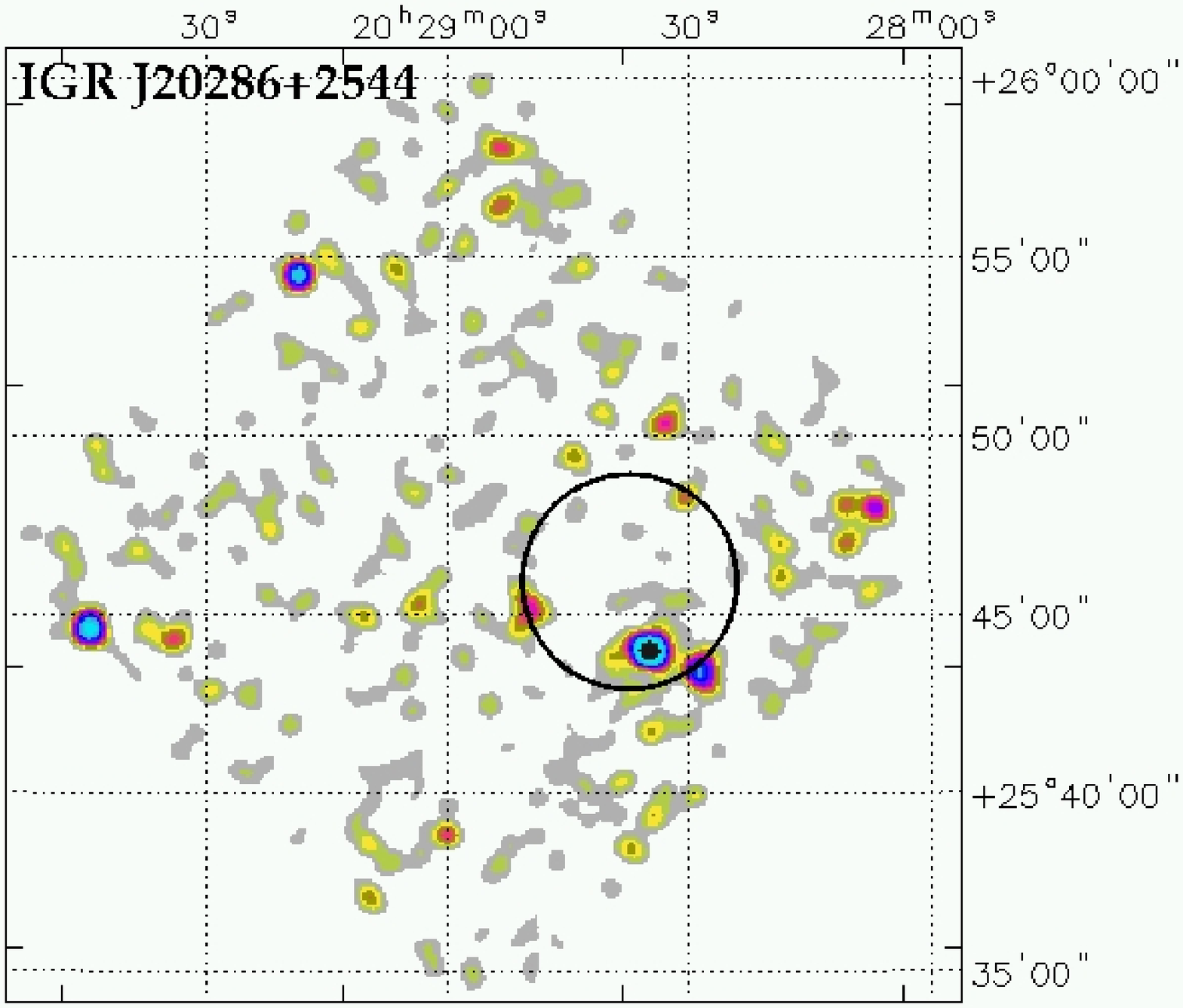,width=5.5cm}}%}
\hspace{-1.0cm}
%\centering{
\mbox{\psfig{file=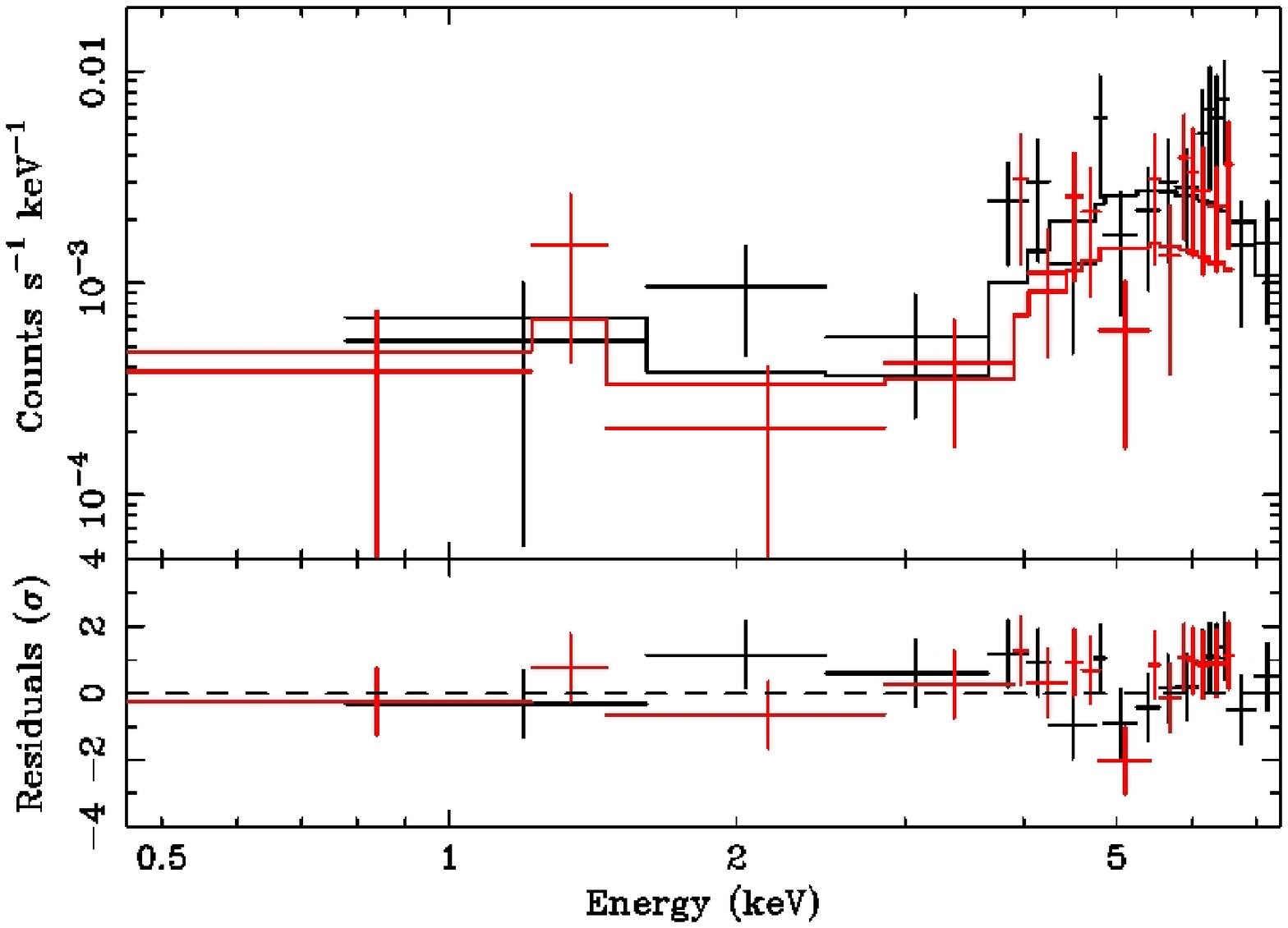,width=5.5cm}}%} 
\caption{XRT 0.3-10 keV image with the associated INTEGRAL error box (black circle) and the corresponding XRT 
spectrum of all the 8 AGN presented. For IGR J14492.5535, the error boxes, corresponding to the INTEGRAL
position as given by \citep{ba06} (black continuum circle) and \citep{re06}(black dashed circle)
are plotted. The images show that for each ISGRI source an X-ray counterpart is always identified.}
\end{center}
\end{figure*}

\section{Conclusions}
 The use of \emph{Swift}/XRTdata has allowed us to characterize the X-ray emission of eight sources
presented here: for four objects the X-ray data confirm their optical classification as Seyfert 1 
galaxies since no intrinsic absorption is found and the spectra are typical of AGN.
For IGR J20286+2544 the intrinsic column density is compatible with a type 2 galaxy, 
confirming again the optical classification. It is likely that this is a Compton thick AGN (\citep{p5}),
as also suggested by the presence of an excess at around 6.4 keV modelled with 
a narrow iron line having an equivalent width (EW) of about 700 eV. Two sources of this sample are still
not optically classified (IGR J14492-3036 and IGR J20187+4041) and they are both absorbed in X-rays, suggesting
a type 2 classification.
IGR J12415-5750 ia a peculiar source as it is a type 2 Seyfert galaxy showing no intrinsic absorption.

\section{Acknowledgements}
This research was supported by ASI under
contract I/R/046/04 and I/R/023/05. This research has made use of data obtained from NED (Jet Propulsion
Laboratory, California Institute of Technology), SIMBAD 
(CDS, Strasbourg, France) and HEASARC (NASA's Goddard Space Flight Center).

\end{document}